# Extreme ultraviolet transient gratings: A tool for nanoscale photoacoustics

L. Foglia[1,*,†], R. Mincigrucci[1,†], A. A. Maznev[2], G. Baldi[3], F. Capotondi[1], F. Caporaletti[4,5], R. Comin[6], D. De Angelis[1], R. A. Duncan[2], D. Fainozzi[1], G. Kurdi[1], J. Li[6], A. Martinelli[7], C. Masciovecchio[1], G. Monaco[7], A. Milloch[8], K. A. Nelson[2], C. A. Occhialini[6], M. Pancaldi[1,9], E. Pedersoli[1], J. S. Pelli-Cresi[10], A. Simoncig[1], F. Travasso[11,12], B. Wehinger[1,13,‡], M. Zanatta[3], F. Bencivenga[1]

1. Elettra - Sincrotrone Trieste S.C.p.A., 34149 Basovizza, Trieste, Italy
2. Department of Chemistry, Massachusetts Institute of Technology, Cambridge, Massachusetts 02139, USA
3. Department of Physics, University of Trento, Povo, Trento I-38123, Italy
4. Van der Waals-Zeeman Institute, Institute of Physics, University of Amsterdam, 1098XH Amsterdam, The Netherlands
5. Van 't Hoff Institute for Molecular Sciences, University of Amsterdam, Science Park 904, 1098XH Amsterdam, The Netherlands
6. Department of Physics, Massachusetts Institute of Technology, Cambridge, Massachusetts 02139, USA
7. Department of Physics and Astronomy, Università di Padova, 35131 Padova, Italy
8. Department of Mathematics and Physics, Università Cattolica del Sacro Cuore, Brescia, I-25133, Italy
9. Department of Molecular Sciences and Nanosystems, Ca' Foscari University of Venice, 30172 Venezia, Italy
10. Istituto Italiano di Tecnologia, via Morego 30, 16163 Genoa, Italy
11. Università di Camerino, 62032 Camerino, Italy
12. INFN, Sezione di Perugia, 06123 Perugia, Italy
13. Department of Molecular Sciences and Nanosystems, Ca' Foscari University of Venice, 30172,400 Venezia Mestre, Italy

*Abstract:*

Collective lattice dynamics determine essential aspects of condensed matter, such as elastic and thermal properties. These exhibit strong dependence on the length-scale, reflecting the marked wavevector dependence of lattice excitations. The extreme ultraviolet transient grating (EUV TG) approach has demonstrated the potential of accessing a wavevector range corresponding to the 10s of nm length-scale, representing a spatial scale of the highest relevance for fundamental physics and forefront technology, previously inaccessible by optical TG and other inelastic scattering methods. In this manuscript we report on the capabilities of this technique in the context of probing thermoelastic properties of matter, both in the bulk and at the surface, as well as discussing future developments and practical considerations.

## 1  Introduction:

Material properties such as elasticity, thermal conductivity or heat capacity are mostly determined by collective lattice dynamics, which exhibit strong length-scale dependencies. Therefore, the thermoelastic response of a system can become drastically different when the spatial dimensions reduce from macroscopic to microscopic scales, i.e. to sizes comparable with interatomic distances or the characteristic length-scales of nanostructures. As an example, thermal transport mechanisms depend on the relative size between the heat source and the mean free path distribution of heat carriers [1–4]. In non-metallic solids, where heat is primarily carried by phonons, its transport is diffusive and follows the Fourier's law for characteristic dimensions much larger than the phonon mean free path, while in the opposite limit phonons move ballistically without collisions. Both descriptions, however, break down when the source size is comparable to the phonon mean free path [1–4]. For instance, on approaching this regime, the heat diffusion of crystalline silicon substantially differs from kinetic expectations at macroscopic length-scales [3] and even becomes overestimated by one order of magnitude at nanometer scales [4]. Another seminal example is given by the thermal and elastic properties of amorphous solids, which at low temperatures drastically differ from those of their crystalline counterparts and, more importantly, are remarkably similar to each other [5]. These deviations from the Debye model that characterize systems without translational symmetries have been attributed to the so-called "boson peak", i.e. a significant excess of vibrational modes in the THz regime [6–8]. Again, collective dynamics of amorphous solids can be described easily in the macroscopic limit, where the system is approximated as a continuum, and in the single-particle limit, but theory fails in the description of the mesoscopic regime, for length-scales comparable to the topological disorder [9–11].

The measurement, and thus the understanding, of the thermoelastic response of matter at the nanoscale is paramount for technological applications exploiting either heat or vibrations, such as phonon engineering in hetero- [12,13] and confined [14] structures, thermal barrier coatings [15], heat assisted magnetic recording [16], nano-enhanced photovoltaics, thermoelectric energy conversion, high power optoelectronics, etc.

---

* Corresponding Author. Email address: laura.foglia@elettra.eu
† The two authors contributed equally.
‡ Present address: European Synchrotron Radiation Facility, 71 Avenue des Martyrs, Grenoble 38000, France

Over the years, an obstacle to the full description of thermoelastic responses in this regime was given by the lack of experimental techniques capable of accessing such a spatial scale [17]. Collective lattice dynamics in condensed matter can be measured, e.g., by inelastic scattering experiments such as inelastic x-ray and neutron scattering for exchanged momentum $Q > 1$ nm$^{-1}$, or Brillouin and Raman scattering for $Q < 0.1$ nm$^{-1}$. The intermediate $Q = 0.1 - 1$ nm$^{-1}$ is hardly accessible by these techniques, despite efforts that have been made for extending Brillouin spectroscopy to the UV range [18] and for improving the performance of x-ray spectrometers [19]. In addition, spectroscopic methods are inherently limited by the instrumental resolution when measuring narrow lines, i.e. long dynamics. This limitation does not affect time-domain techniques based on lasers, such as picoacoustics or transient grating (TG) spectroscopy, which are however limited to small $Q$ by the relatively long optical wavelength. Despite attempts to overcome this [20], to date no optical method is capable to access the intermediate $Q$ regime.

The advent of free electron laser (FEL) sources has offered EUV pulses of sufficient brilliance to allow for the extension of non-linear optical techniques to shorter wavelengths, among others TG spectroscopy [21–23]. The EUV TG approach has been pioneered at the FERMI FEL (Trieste, Italy) with the dedicated endstation TIMER [24]. First results beyond the proof of principle [25] reveal that EUV TG is capable of incisively and selectively studying bulk and surface phonons [4,26], thermal transport kinetics [14,27] and magnetic dynamics [28–30]. The recent and steady development of the technique has evidenced specific peculiarities with respect to the well-established optical TG spectroscopy that need to be considered both when predicting as well as when interpreting experimental EUV TG results. In this paper, we provide a general overview of the TG approach (section 2) and a discussion on its extension to the EUV (section 3), with a short description of the experimental setup at FERMI. We then present some exemplary experimental results showing how the technique can be exploited to investigate nanoscale thermoelastic responses in several classes of materials (section 4) and finally briefly discuss future developments (section 5).

## 2 The transient grating approach

TG is a third order (four-wave-mixing) non-linear optical technique, where three optical fields interact with the sample to generate a fourth (signal) field, as depicted in figure 1(a). In particular, two light pulses of equal wavelength λ (called *pumps*) are overlapped in time and space on the sample with a crossing angle 2θ. The interference of these two pulses generates a spatially periodic excitation, which results in a modulation of light intensity or polarization, depending if the pump pulses have respectively parallel or orthogonal polarizations [31].

Assuming that the sample is a slab with the surface oriented orthogonally to the bisector of the pump pulses (cfr. figure 1(a)), the spatial periodicity of the sinusoidal excitation pattern $\Lambda_{TG}$ depends only on λ and 2θ as:

$$\Lambda_{TG} = \lambda/[2\sin(\theta)], \qquad (Eq.\ 1)$$

and the wavevector of the TG excitation, $\boldsymbol{Q}_{ex} = \pm\boldsymbol{Q}_{TG}$ (with $Q_{TG} = 2\pi/\Lambda_{TG}$), is parallel to the sample surface and lies in the plane defined by the two pump beams. Such a patterned excitation effectively acts as a transient diffraction grating for a third variably-delayed pulse of wavelength $\lambda_{pr}$ and wavevector $\mathbf{k}_{pr}$, the *probe*, giving rise to a fourth pulse: the diffracted beam, or *signal*, that in typical TG experiments has the same wavelength as the probe and wavevector $\mathbf{k}_{sig}$. The signal beam parameters (intensity, polarization, etc.) as a function of time delay Δt encode the dynamics of the photoexcited processes that are characterized by the wavevector $\boldsymbol{Q}_{ex}$. Eq. 1 also sets a lower bound at λ/2 for $\Lambda_{TG}$, corresponding to $Q_{ex} < 4\pi/\lambda$.

In the considered geometry and with the further assumption that the grating thickness d is much smaller than both $\Lambda_{TG}$ and $\lambda_{pr}$, the conditions for observing a diffracted signal are given by the thin grating equation [32]:

$$\Lambda_{TG}[\sin(\theta_{sig}) \pm \sin(\theta_{pr})] = \pm m\lambda_{pr}, \qquad (Eq.\ 2)$$

where $\theta_{pr}$ and $\theta_{sig}$ are respectively the incidence angle of the probe and the diffraction angle of the signal, and $\pm m$ accounts for the diffraction order, following the convention depicted in figure 1(b). Consequently, Eq. 2 defines an upper limit for the probe wavelength of $\lambda_{pr} < 2\Lambda_{TG}$ in order to observe a signal.

When the thin grating condition d << $\Lambda_{TG}$ is not fulfilled, the diffraction efficiency in forward diffraction $\eta_F$ also depends on the wavevector mismatch, $\Delta Q_z = 2\pi n(\lambda_{pr})|\cos\theta_{pr} - \cos\theta_{sig}|/\lambda_{pr} = |\cos\theta_{pr} - \cos\theta_{sig}|n(\lambda_{pr})k_{pr}$, which is determined by the energy conservation in the diffraction process (i.e. $|k_{pr}|=|k_{sig}|$) and by the continuity of the tangential component $\boldsymbol{Q}_x$ of the electric field at the vacuum-sample interface (see figure 1(c)):

$$\eta_F = \frac{I_{sig}}{I_{pr}} = \frac{\cos\theta_{sig}}{\cos\theta_{pr}}\frac{|E_{sig}|^2}{|E_{pr}|^2} = \frac{\cos\theta_{sig}}{\cos\theta_{pr}}|\Delta n(\lambda_{pr})|^2[\pi n(\lambda_{pr})d/\lambda_{pr}\cos(\theta_{pr})]^2[\sin(\Delta Q_z d/2)/(\Delta Q_z d/2)]^2, \qquad (Eq.\ 3)$$

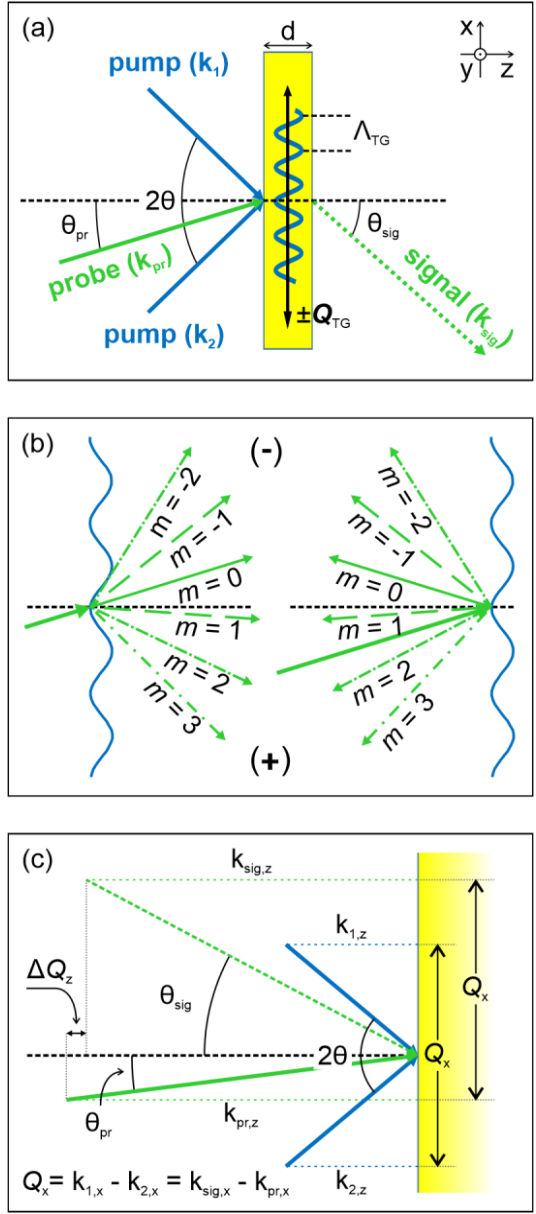

*Figure 1: (a) Schematic of the TG experiment evidencing the relevant experimental parameters: wavevectors of incoming and signal beams ($k_1$, $k_2$, $k_{pr}$ and $k_{sig}$), pump crossing angle ($2\theta$), probe incidence angle $\theta_{pr}$, signal diffraction angle $\theta_{sig}$, transient grating wavevector $Q_{TG}$ and period $\Lambda_{TG}$, and grating (or sample) thickness d. The reference frame is indicated in the top-right corner. (b) Definition of the sign of the diffraction orders in forward (left sketch) and backward (right sketch) diffraction geometry. (c) Continuity of the tangential component of the electric fields $Q_x = k_{1,x}+k_{2,x} = k_{pr,x}+k_{sig,x}$, and wavevector mismatch ($\Delta Q_z$).*

where $I_{pr}$ and $I_{sig}$ are the pulse energies of the transmitted probe beam and of the 1st diffraction order, respectively, $n(\lambda_{pr})$ the refractive index in the medium at the probe wavelength, and $|\Delta n(\lambda_{pr})|$ the amplitude of its variation; note that the $\sin(\Delta Q_z d/2)/(\Delta Q_z d/2)$ term tends to unity either for $d \to 0$, as in the case of surface excitations, or for $\Delta Q_z \to 0$. The latter limit is reached when the so-called Bragg condition, i.e. $\theta_{pr} = \theta_{sig} = \theta_B$, is met. In this case the TG signal can build up in the bulk irrespective of the value of d (volume grating diffraction) and is therefore greatly enhanced. For finite values of d the TG efficiency is significant only for $\Delta Q_z d/2 \ll 1$. Eq. 3 assumes: i) a uniform excitation profile along z, i.e. a negligible pump absorption, ii) a negligible probe absorption, iii) infinite spatial extension of pump and probe beams, iv) weak diffraction ($\eta_F \ll 1$), i.e. the energy transferred from the three input fields to the output (diffracted) one is a negligible fraction of the total energy, and v) a negligible contribution from surface displacement or thickness modulations. As outlined in the following, some of these assumptions do not hold for EUV TG and Eq. 3 has to be modified accordingly.

## 3 Extension of TG spectroscopy to the EUV regime

The discussion following Eq. 1 clearly shows that TG can access larger Q's by using shorter pump and probe wavelengths, i.e. entering the EUV/x-ray regime. The first attempts in this direction have employed high harmonic generation (HHG) sources to probe optically-excited TGs. While HHG-based TG is still limited in $Q_{ex}$ by the optical wavelength of the pump, probing in the EUV regime introduces the capability of exploiting core resonances and, thus, to be element specific. This allowed to understand the ultrafast mechanism of nonlinear signal generation in atomic helium [33] and to study the insulator to metal transition in crystalline VO$_2$ [34]. Additionally, the shorter wavelength and penetration depth of EUV light makes the HHG probe particularly sensitive to surface excitations, such as surface acoustic waves (SAWs) [35], and enables studying the thermoelastic response of nano-patterned surfaces [36,37].

However, the brilliance of HHG sources used in these pioneering experiments was too low to exploit them for the grating generation. While in certain cases (e.g. for relatively long EUV excitation wavelengths) state of the art HHG sources may be sufficient to stimulate EUV TGs [38], only the advent of FEL sources has led to the potential of routinely generating EUV and x-ray pulses of sufficient brilliance to allow for the excitation of non-linear optical processes, including TG [23,39–42]. Using high-energy photons to drive the TG excitation introduces some peculiarities with respect to the optical excitation, that go beyond the mere reduction of $\Lambda_{TG}$ and that are the focus of this paper.

### 3.1 EUV excitation

In the optical regime the excitation mechanisms change drastically depending on the investigated kind of materials, i.e. if we are considering a weakly absorbing dielectric medium, a semiconductor or a metal, and range from field-driven density modulations (electrostriction) to intensity-driven temperature and electronic population gratings. Moreover, optical photons are essentially not transmitted into metals due to their frequency being generally lower than most plasma frequencies and thereby making optically-excited TG studies of bulk metals impossible.

For EUV pump the situation is drastically different: their photon energy in the 10s to 100s eV range is always larger than plasma frequencies in all kind of materials, as well as larger than typical electronic band gaps, making the distinction between dielectrics, semiconductors and metals irrelevant in the excitation process.

This holds analogously for the probe frequency and, assuming that it is also far away from core resonances, one can apply the approximation for the refractive index that is commonly used for x-rays, i.e.: $n = 1 - \delta + i\beta$, where both $\delta$ and $\beta$ depend linearly on the total electron density [43].

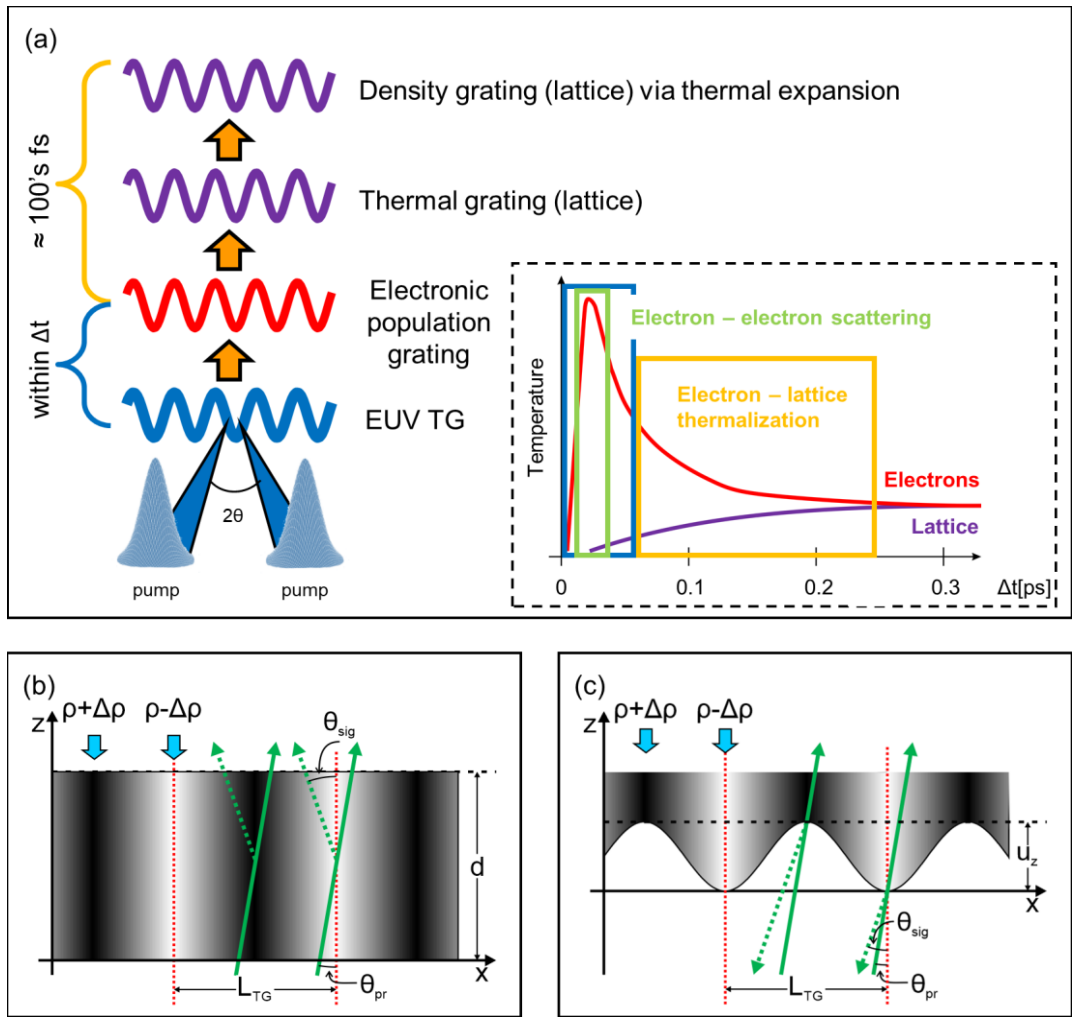

*Figure 2: (a) Main excitation mechanism for EUV TG: a population grating of electronic excited states is generated "instantaneously", in a 100's fs timescale it thermalizes with the lattice leading to temperature and subsequent density gratings, as sketched in the inset for the case of temperature. Here the blue and orange boxes represent, respectively, the time duration of the excitation pulse and the typical timescale for the electronic-lattice equilibration. The green box encloses the timescale of electron-electron scattering. Panels (b) and (c) sketch, respectively, the leading contributions to forward (bulk density modulation) and backward diffraction (surface displacement); darker areas represents denser (cold) regions of the sample, lighter ones the hotter regions, solid and dashed lines schematize the probe and signal beams, respectively, while $u_z$ is the surface displacement.*

The dominating excitation mechanism in the EUV regime is the generation of a photoexcited electronic population with very high excess energy, which is exchanged with the surrounding electron bath within a few tens of femtoseconds and without a substantial diffusion on the length-scales of $\Lambda_{TG}$. Since such dynamics are much faster than the time duration of the FEL pulses ($\Delta t_{FEL} \approx 50$ fs) so far employed in EUV TG experiments, they can be regarded as "instantaneous" and lead to stripes of hotter electrons alternating with stripes of cold ones with the same periodicity $\Lambda_{TG}$ as the TG (population grating). This modulated hot electron population thermalizes with the lattice via electron-phonon scattering, resulting in the formation of a thermal grating within hundreds of fs, as depicted in figure 2(a). Thermal expansion leads then to an alternation of cold sample regions with increased density $\rho + \Delta\rho$ (the unexcited stripes) and hot ones with lower density $\rho - \Delta\rho$ (figure 2(b)), being $\rho$ the density of the unperturbed sample. Similarly, at the surface, photoexcitation and the consequent thermal expansion will lead to a periodic surface displacement, with a peak to valley distance between hot and cold areas of $u_z$ (figure 2(c)). Bulk and surface expansion will launch respectively longitudinal (LA) and surface acoustic waves with a wavevector $Q_{ex}$.

The incident probe and the signal beam are schematized in both figure 2(b) and 2(c) as two rays (solid green arrows) passing through a high- and a low-density region excited by the EUV TG. In forward diffraction geometry (figure 2(b)) the diffracted signal is related mostly to the phase difference accumulated by these two rays while they propagate through the sample. In backward diffraction (figure 2(c)) the signal is instead mainly given by the optical path difference between the rays reflected by the peaks and the valleys of the surface wave. Nevertheless, surface modulation may also contribute to the peak-null differences of forward diffraction and the differences in refractive index may affect the reflectivity of the sample, altering the backward diffraction efficiency.

**EUV TG signal in forward diffraction geometry and the role of absorption**

In order to estimate $\eta_F$ in an EUV TG experiment, one should evaluate the effect on $|\Delta n(\lambda_{pr})|^2$ of the dominating EUV TG excitation mechanism, sketched in Figure 2 ("instantaneous" electronic population grating, ultrafast relaxation into a lattice temperature grating with the contextual formation of a density grating via thermal expansion). $|\Delta n(\lambda_{pr})|^2$ is thus expected to show a component due to the modulation of the excited electron density, one due to the temperature variation and one associated to the density changes. Assuming that within the few fs time-scale of electron thermalization there is no relevant transverse electron diffusion on the length-scale of $\Lambda_{TG}$, which could reduce the contrast, for an EUV probe far from any core resonances the dominant component to $|\Delta n(\lambda_{pr})|^2$ is given by the density variation ($\Delta\rho$), i.e.:

$$|\Delta n(\lambda_{pr})| \approx \left|\frac{\partial n(\lambda_{pr})}{\partial \rho}\Delta\rho\right| = (\Delta\rho/\rho)\sqrt{\delta^2 + \beta^2} \approx \alpha_v \Delta T \sqrt{\delta^2 + \beta^2}, \qquad \text{(Eq. 4)}$$

where we considered $n = 1 - \delta + i\beta$ and a linear dependence of both $\delta$ and $\beta$ on $\rho$ [43], as mentioned above, and that the changes in ρ are predominantly associated to a temperature variation (ΔT) via volumetric thermal expansion ($\alpha_v$), i.e.: $\Delta\rho/\rho = \alpha_v \Delta T$. The magnitude of ΔT can be roughly estimated from the FEL excitation fluence (F), sample's specific heat (c) and effective excitation length in the sample (L) as $\Delta T = F/(\rho c L)$, where L is the smaller value between $L_{abs}$ and d. As an order of magnitude estimate one may consider reasonable values as $\alpha_v = 10^{-5}$ K$^{-1}$, $\delta = \beta = 10^{-2}$ and ΔT = 80 K (e.g. from F = 33 mJ/cm$^2$, $\rho$ = 3.4 g/cm$^3$, c = 0.7 J g$^{-1}$ K$^{-1}$, L = 100 nm), resulting in $|\Delta n(\lambda_{pr})|^2 \approx 10^{-10}$.

Furthermore, Eq. 3 has to be modified to account for some of the assumptions that no longer hold in the EUV regime. In particular, in most materials, if not all, the absorption length of EUV light is on the order of a few tens of wavelengths at most. Therefore, the signal loss due to absorption prevails over the bulk signal enhancement for sufficiently thick samples and Eq. 3 needs to be modified to account for absorption in all pump, probe and signal fields: a situation rarely considered in optical TG experiments. Assuming an exponential decay of the pump and probe fields along z, with decay length $2L_{abs}\cos\theta$ and $2L_{abs,pr}\cos\theta_{pr}$, respectively, and a factor $e^{-(d-z)/2L_{abs,pr}\cos\theta_{sig}}$ to account for the absorption of the signal field into the material, the forward diffraction efficiency from a slab-shaped sample with thickness $d$ can be approximated as:

$$\eta_F = \frac{\cos\theta_{sig}}{\sin\theta_{pr}}|\Delta n(\lambda_{pr})|^2 \left[\pi n d(\lambda_{pr})/\lambda_{pr}\cos\theta_{pr}\right]^2 \frac{e^{-d/L^*} - 2\cos(\Delta Q_z d)e^{-d/2L^*} + 1}{(d/2L^*)^2 + (\Delta Q_z d)^2} e^{-d/L_{abs,pr}\cos\theta_{sig}}, \qquad \text{(Eq. 5)}$$

where $L^* = \left((L_{abs}\cos\theta/2)^{-1} + (L_{abs,pr}\cos\theta_{pr})^{-1} - (L_{abs,pr}\cos\theta_{sig})^{-1}\right)^{-1}$. For $L^* \to \infty$, which is the case when absorption can be neglected, the fraction reduces to the $[\sin(\Delta Q_z d/2)/(\Delta Q_z d/2)]^2$ term of Eq. 3. On the other hand, the limit $L^* \to 0$ can be reached either when $L_{abs} \to 0$ or $L_{abs,pr} \to 0$. In the latter case $\eta_F \to 0$ because the signal is no longer transmitted through the sample ($e^{-d/L_{abs,pr}\cos\theta_{sig}} \to 0$). In the former case the dependency on both d and $\Delta Q_z$ drops, except the dependence on d of the probe and signal absorption, resulting in $\eta_F \to 0$ proportionally to the factor $(L_{abs}/\lambda_{pr})^2 \to 0$. This allows to obtain a relatively simple analytical expression that generalizes Eq. 3 to the case of finite absorption lengths.

Figure 3 displays some representative plots obtained from Eq. 5, to illustrate the effects of absorption in both pump and probe as well as those of $\Delta Q_z$, as a function of sample thickness d for a given value of $|\Delta n(\lambda_{pr})|^2$. In all plots we considered a set of parameters that match the Bragg condition and are typically used in EUV TG experiments at TIMER [24]: $\lambda$ = 39.9 nm, $\lambda_{pr}$ = 13.3 nm, $\theta$ = 13.8° and $\theta_{pr}$ = 4.6°. The solid black curve in all panels of figure 3 represents the quadratic dependence of $\eta_F$ on d resulting from Eq. 3 and from Eq. 5 when $L_{abs} \to \infty$, $L_{abs,pr} \to \infty$ and $\Delta Q_z \to 0$. The effect of pump absorption only is to limit the signal increase $\propto d^2$, which reaches a constant level for $d \gg L_{abs}$, as displayed in figure 3(a); here realistic values for EUV excitation ($L_{abs}$ = 200/100/50 nm) are considered respectively in the solid, dashed and dotted red curves. The effect of the probe absorption only is shown in blue in figure 3(b) for

$L_{abs,pr} = 200/100/50$ nm (again as solid, dashed and dotted curves respectively) and essentially results in an exponential decrease of the signal. Since the d-dependence is stronger for the exponential decrease than for the quadratic growth, this results in $\eta_F \to 0$ for $d \gg L_{abs,pr}$ and, consequently, in an optimal sample length, i.e. a maximum in $\eta_F$ vs d at a finite value of $d=d^*$. We note that $d^* = 2L_{abs,pr}$ when the probe absorption is the only sizable effect in Eq. 5, a condition reached for $L_{abs} \to \infty$ and $\Delta Q_z \to 0$. The combined effect of both pump and probe absorption is reported in figure 3(c), where the orange traces show the result obtained respectively for $L_{abs} = L_{abs,pr} = 200$ nm (solid), $L_{abs} = L_{abs,pr} = 100$ nm (dashed) and $L_{abs} = L_{abs,pr} = 50$ nm (dotted). This represents the typical situation of EUV TG experiments and determines both a large decrease in $\eta_F$ and a value of $d^*$ substantially lower than $2L_{abs,pr}$. The aforementioned order of magnitude estimate $|\Delta n(\lambda_{pr})|^2 \approx 10^{-10}$ results in $\eta_F \approx 10^{-8}$, which can yield to a tangible signal if one considers that typically FEL-based experimental setups can deliver EUV pulses with more than $10^8$ photons at the sample.

In the context of typical parameter ranges for EUV TG, the effect of $\Delta Q_z$ is less relevant than the one of absorption, as evidenced by the comparison of the orange lines in figure 3 (c) with the corresponding green ones. They are calculated from Eq. 5 assuming, on top of the absorption, a wavevector mismatch of $\Delta Q_z = 0.02 * k_{pr}$, a representative value of the bandwidth ($\Delta\lambda/\lambda \approx 2$ %) of most FEL sources. Figure 3(d) shows the effect of $\Delta Q_z$ only, as obtained by considering $L_{abs} = L_{abs,pr} \to \infty$. Even more, for the narrowband ($\Delta\lambda/\lambda < 0.1$ %) case of FERMI, where most EUV TG experiments were performed to date, $\Delta Q_z$ does not introduce any significant effect on $\eta_F$ in the Bragg geometry considered here. However, $\Delta Q_z$ can become relevant regardless of the source bandwidth when Bragg conditions are not satisfied.

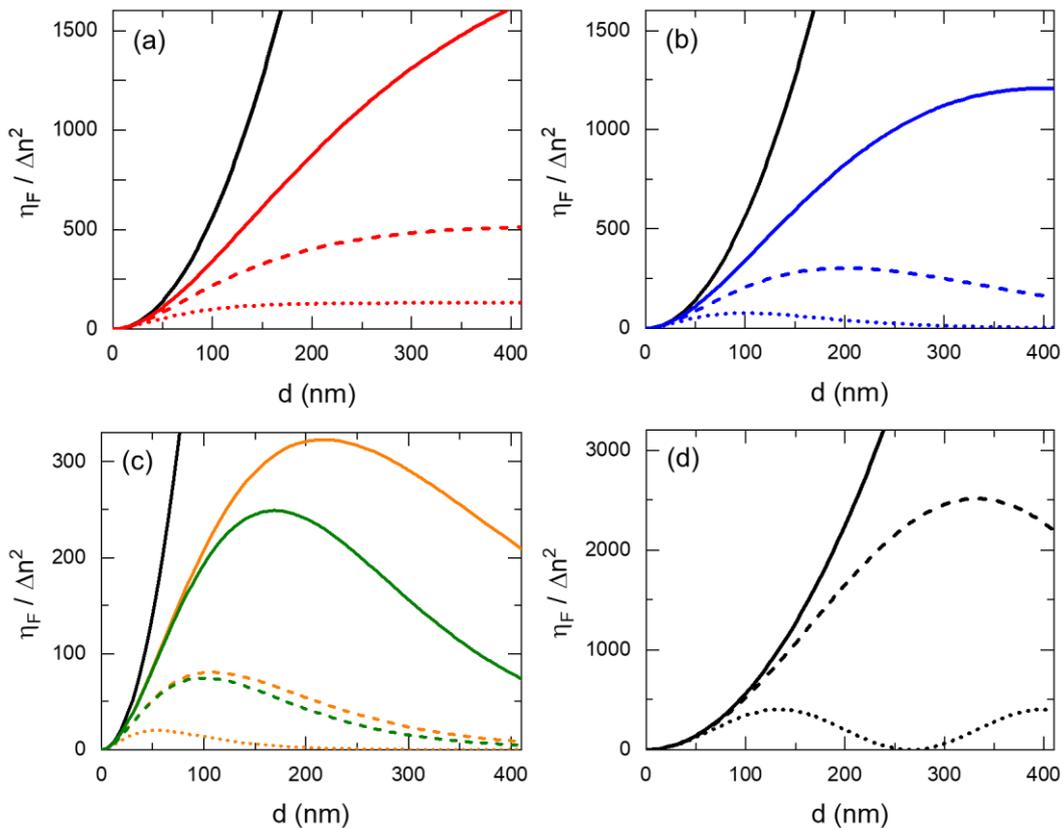

*Figure 3: (a) Dependence of $\eta_F/\Delta n^2$ on d for $L_{abs,pr} \to \infty$, $\Delta Q_z \to 0$ and $L_{abs} = 200$ nm (solid red curve), 100 nm (dashed red curve) and 50 nm (dotted red curve). (b) Dependence of $\eta_F/\Delta n^2$ on d for $L_{abs} \to \infty$, $\Delta Q_z \to 0$ and $L_{abs,pr} = 200$ nm (solid blue curve), 100 nm (dashed blue curve) and 50 nm (dotted blue curve); note the maximum reached at $d = d^* = 2L_{abs,pr}$. (c) Dependence of $\eta_F/\Delta n^2$ on d for $\Delta Q_z \to 0$ and $L_{abs} = L_{abs,pr} = 200$ nm (solid orange curve), 100 nm (dashed orange curve) and 50 nm (dotted orange curve). The solid and dashed green curves are calculated by considering $\Delta Q_z/k_{pr} = 0.02$ for $L_{abs} = L_{abs,pr} = 200$ nm and 100 nm, respectively; note the reduced vertical scale. (d) Dependence of $\eta_F/\Delta n^2$ on d for $L_{abs} = L_{abs,pr} \to \infty$ and $\Delta Q_z/k_{pr} = 0.02$ (dashed curve) and 0.05 (dotted curve); note the expanded vertical scale. The solid black curve in all panels is the $d^2$ dependence of $\eta_F$, obtained for $L_{abs} \to \infty$, $L_{abs,pr} \to \infty$ and $\Delta Q_z \to 0$. See text for further details on the other parameters used in the calculations.*

Figure 4 illustrates the dependence of $\eta_F$ (normalized to its maximum value: $\eta_F^*$) on $\Delta Q_z/k_{pr}$ in a "thick" (d=2 µm; black lines) and a "thin" sample (d=200 nm; red traces), both neglecting or considering the pump's absorption. These plots are calculated from Eq. 5 by considering the aforementioned (Bragg) conditions, i.e: $\lambda = 39.9$ nm, $\lambda_{pr} = 13.3$ nm, $\theta = 13.8°$ and $\theta_{pr} = 4.6°$, with the further assumption $L_{abs,pr} \to \infty$. One can notice how the effect of $L_{abs}$ is analogous to a reduction of d, and results in a substantial broadening in the $\eta_F/\eta_F^*$ vs $\Delta Q_z/k_{pr}$ curve. On the other hand, the probe's absorption (not considered in figure 4(a)) has no substantial effects in the shape of this curve, while it may have large effects on the magnitude of $\eta_F$ (see figure 3(b)). The possibility to have tangible values of $\eta_F$ far from ideal Bragg conditions often allows to simplify the experimental setup. For example, in figure 4(b) we plot the diffraction efficiency as a function of $\theta_{pr}$ for two conditions considered in figure 4(a), i.e. a "thick" sample with no absorption (black) and an absorbing "thin" sample (dashed red); the latter is the realistic case for EUV TG. One can readily notice how here there is no need to precisely control $\theta_{pr}$, and thus respect the Bragg condition. Analogously, an appreciable signal can be detected by varying $\Lambda_{TG}$ in quite a large range for a given $\theta_{pr}$, as displayed in figure 4(c). Here, the curves were calculated varying λ from 70 to 7 nm and computing $\Lambda_{TG}$ via Eq. 1. A better estimate for a given sample can be provided by using a λ-dependent value for $L_{abs}$. Moreover, we recall that all curves in figure 4 do not account for the probe's absorption, which is important to determine the optimal sample thickness and to estimate the absolute value of $\eta_F$. In the case of figures 4(b) and 4(c), the probe absorption may change the shape of the curves, since either a variation in $\theta_{pr}$ or in $\Lambda_{TG}$ leads to a variation in $\theta_{sig}$, according to Eq. 2, with a consequent effect on the signal attenuation.

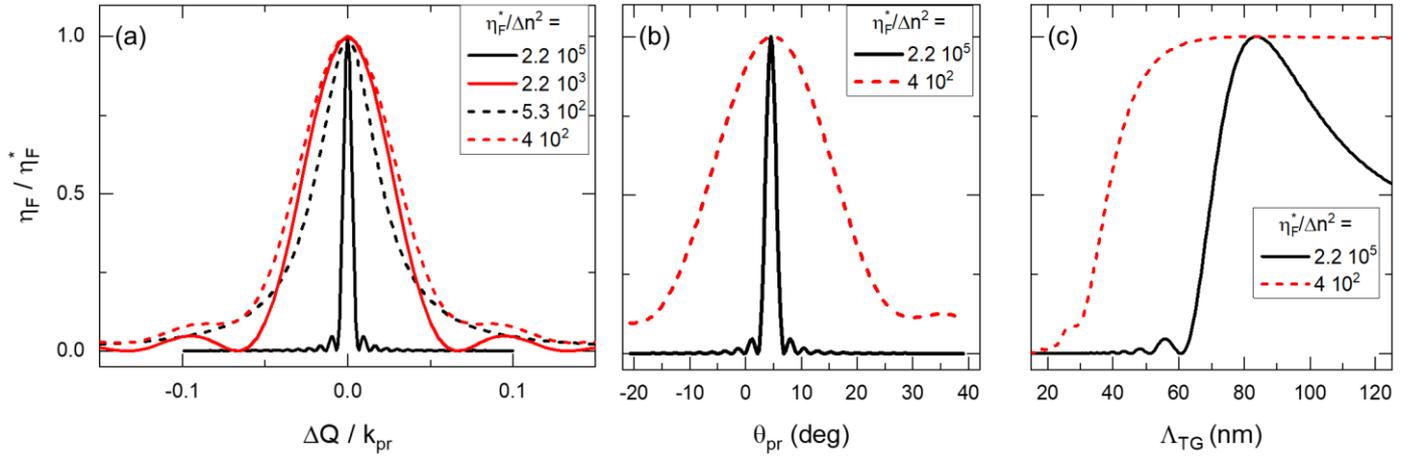

Figure 4: (a) Dependence of $\eta_F/\eta_F^*$ on $\Delta Q_z/k_{pr}$ (where $\eta_F^*$ is the maximum value of $\eta_F$) in a thick (d=2 µm; black lines) and a thin (d=200 nm; red lines) sample, calculated by considering both $L_{abs} \to \infty$ (solid lines) and $L_{abs}$ =100 nm (dashed lines). (b) Dependence of $\eta_F/\eta_F^*$ on $\theta_{pr}$ for d=2 µm and $L_{abs} \to \infty$ (solid black line) and for d=200 nm and $L_{abs}$ =100 nm (dashed red line); the Bragg angle is $\theta_{pr}$ =4.6°. (c) Dependence of $\eta_F/\eta_F^*$ on $\Lambda_{TG}$ for d=2 µm and $L_{abs} \to \infty$ (solid black line) and for d=200 nm and $L_{abs}$ =100 nm (dashed red line); the Bragg conditions are meet for $\Lambda_{TG}$ = 83.6 nm. The values of $\eta_F^*$ are reported in the legend of the individual panels.

We stress that the behavior illustrated in Figure 3 and 4 is typical in the context of using EUV light in condensed matter, but it is not necessarily restricted to this spectral range, since it arises from general aspects of the TG process. For instance, the same behavior is expected in optical TG experiments from samples showing absorption lengths comparable to the laser wavelengths, such as semiconductors with resonant electronic transitions. In realistic 3.3 conditions, despite some advantages related to the tolerance to large values of $\Delta Q_z$ (relaxed Bragg conditions), this situation is overall unfavorable, since absorption implies a decrease as large as orders of magnitudes in $\eta_F$, and thus in the experimental signal.

### EUV TG signal in backward diffraction geometry

As mentioned in section 3.1, the periodic surface displacement sketched in figure 2(c) is the main responsible for the backward diffracted TG signal, which diffraction efficiency is given by:

$$\eta_B = R(2\pi u_z/\lambda_{pr})^2 \cos\theta_{pr} \cos\theta_{sig}, \qquad (Eq.\ 6)$$

where R is the reflectivity at $\lambda_{pr}$. Eq. 6 obviously holds for any wavelength: it can be applied to estimate the TG signal in the optical as well in EUV regime. However, in the optical regime refractive index modulations due to electronic excitations typically result in a sizable signal [44], while a tangible electronic contribution to the EUV TG signal is expected only when $\lambda_{pr}$ matches an electronic core resonance. Assuming that the system is excited to the same temperature, i.e. for the same surface displacement $u_z$, the decrease of $\lambda_{pr}$ below the optical range and towards the 10s of nm range represents an advantage, because of the $\lambda_{pr}^{-2}$ dependence of $\eta_B$, as already evidenced in initial optical TG pump/EUV probe experiments [35]. However, the value of R drastically reduces in the EUV range, and is much more sensitive to surface roughness. Figure 5(a) illustrates an example of this trade-off for different samples, assuming $u_z$ = 10 pm (in the same range as found in Ref. [26] for the case of SrTiO3) and typical values of incident and diffraction angles ($\theta_{pr} = \theta_{sig} = 15°$), while the values of R are taken from Ref. [45]. These plots show estimated values of $\eta_B$ in the order of $10^{-6} - 10^{-8}$ for $\lambda_{pr} > 10$ nm, resulting in signals on the same order of magnitude as those in transmission geometry.

Figure 5 also highlights how the efficiency of EUV TG experiments in backward diffraction becomes unfavorable on reducing $\lambda_{pr}$, since the decrease in R is no longer compensated by the shortening of $\lambda_{pr}$. Moreover, specific (material-dependent) core resonances have a drastic effect on R, resulting into the dips highlighted in figure 5(a) and stressing the importance of a proper choice of $\lambda_{pr}$. Another critical parameter is the surface roughness at the nm level, which can reduce R by orders of magnitude, in particular for short values of $\lambda_{pr}$ (see dashed and dotted lines in Figure 5(a); calculated from Refs. [45,46]), indicating the critical need of high-quality sample surfaces. However, even in atomically flat samples, the detrimental effect of both surface roughness and decrease in R makes EUV TG in backward diffraction hardly viable for values of $\lambda_{pr}$ below a few nm (see also figure 5(b)). Most likely, this situation will prevent the extension of backward diffraction TG at x-ray wavelengths, thus practically hampering surface sensitive TG experiments in that regime. On the other hand, x-ray TG in forward diffraction is expected to become much more favorable, because the absorption length, and thus $d^2$, largely increases (see figure 5(b)). Indeed, it is interesting to notice how the factors $Ru_z^2$ in Eq. 6 and $(\delta^2 + \beta^2)(\Delta\rho/\rho)^2 d^2$ in Eq. 5 essentially play the same role. Since $R \propto \delta^2 + \beta^2$, the discriminating factor that makes $\eta_F$ favorable over $\eta_B$ as a function of the photon wavelength, is $\eta_F/\eta_B \propto (\Delta\rho/\rho)^2 (d/u_z)^2$. Both $(\Delta\rho/\rho)^2$ and $u_z^2$ are driven by ΔT, i.e. by the same lattice temperature grating, and therefore do not depend on the photon wavelength, while d strongly depends on the photo absorption, as discussed above.

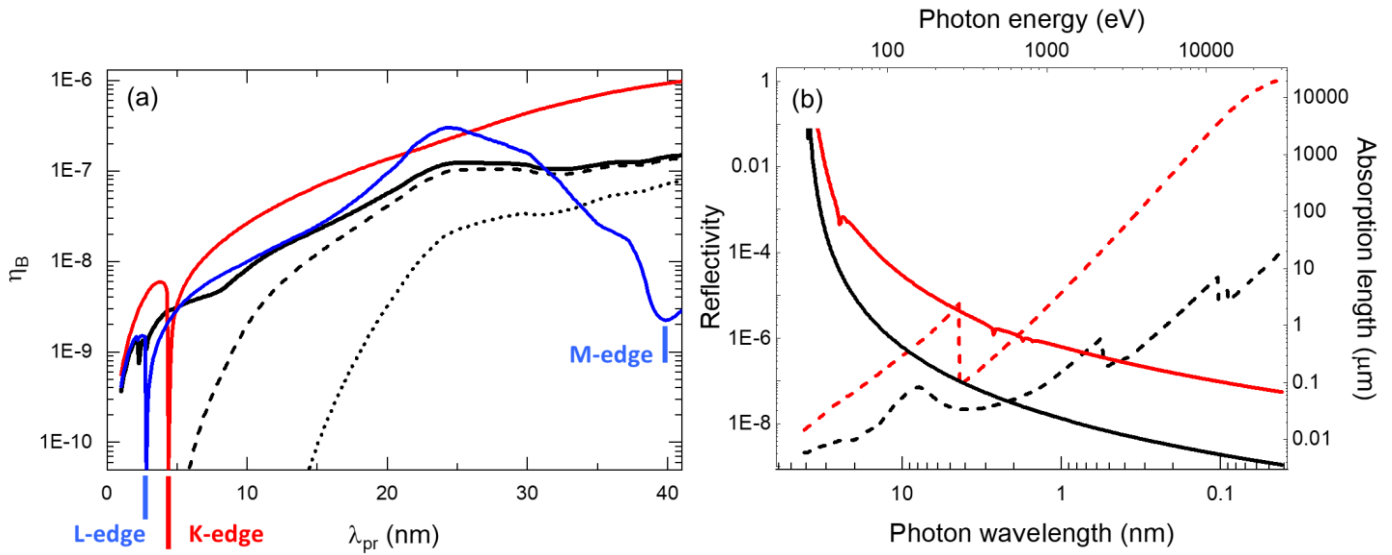

Figure 5: (a) Dependence of $\eta_B$ on $\lambda_{pr}$ displayed in a semi logarithmic scale, calculated from Eq. 6 assuming $u_z = 10$ pm and $\theta_{pr} = \theta_{sig} = 15$ deg. Black, red and blue solid lines respectively refer to different materials: $SrTiO_3$, carbon and titanium; in all these cases we neglected the effect of surface roughness. Dashed and dotted lines correspond to a surface roughness of 1 nm and 3 nm (root-mean-square), respectively, for the case of $SrTiO_3$. The red and blue vertical segments respectively indicate the electronic core resonances of carbon and titanium. (b) Reflectivity (solid lines; left vertical scale) and absorption length (dashed lines; right vertical scale) as a function of photon wavelength in the EUV (< 1 nm) and X-ray range (>1 nm) for two representative elements: C (black lines) and Au (red lines). The photon energy scale is shown on top.

We recall here that a signal originating from surface displacement (or from thickness modulations) can contribute to the EUV TG signal in forward diffraction as well. For a given amplitude of the EUV induced thermal grating, one can estimate the magnitude of this contribution by comparing the optical path difference due to bulk density modulations, $\Delta n^2 = (\delta^2 + \beta^2)(\Delta\rho/\rho)^2 d^2$, with those due to either surface displacements, $\Delta n^2 = u_z^2(\delta^2 + \beta^2)$, or thickness modulations, $\Delta n^2 = \Delta d^2 (\delta^2 + \beta^2)$, where $\Delta d$ is the amplitude of such modulations.

### 3.4 The TIMER instrument

All EUV TG data shown in this work were collected at the TIMER instrument [24] of the FERMI FEL facility (Trieste, Italy), which is able to deliver intense, ultrafast and nearly Fourier transform-limited EUV pulses in the 4-100 nm wavelength range, at a repetition rate of 50 Hz [47,48]. The typical energy per pulse is 10-100 µJ, at the FEL output, while the pulse duration (Δt_FEL) is about 20-70 fs full width at half maximum (FWHM); the shorter the wavelength, the smaller are both the intensity and Δt_FEL. The TIMER instrument is entirely under high vacuum and is able to split the FEL beam in three parts, namely: two almost equal intensity pump beams, with a controllable time delay (in the few ps range) among them, and a weaker probe beam, with a variable time delay (Δt) on the order of 1-2 ns; the exact range (up to 3.5 ns) depends on the specific values of θ and $\theta_{pr}$, while the accuracy (due to the mechanics) is of about one fs. These beams are recombined at the sample position in order to realize the geometry sketched in figure 1(a), with selectable values of θ = 9.2°, 13.8°, 39.5° or 52.7° and $\theta_{pr}$ = 3.05°, 4.56°, 12.2° or 15.5°. The spot size at the sample position is in the 50-300 μm range for both pump and probe; the system is designed to have smaller spot sizes for larger θ. The typical range of pulse energy at the sample for the pump is 0.1-10 μJ (shorter wavelengths and larger θ result in lower intensity), while that of the probe beam is 0.1-10 μJ, and strongly depends on the chosen value of $\lambda_{pr}$ (see Table I). The value of $\lambda_{pr}$ can be selected in narrow bandwidths around 6 possible values: $\lambda_{pr}$ = 20.4 nm, 17.7 nm, 16.6 nm, 13.3 nm, 8.34 nm and 6.71 nm that are determined by special multilayer optics in the probe's delay line. Some of these values of $\lambda_{pr}$ correspond to specific core resonances in selected materials, as indicated in Table I. As discussed in sections 3.1 and 3.2, core resonances are not relevant to address the thermoelastic response, however, they can be exploited to study the electronic and magnetic response, with the additional benefit of element selectivity [28,30]. The pump wavelength is related to $\lambda_{pr}$ as: $\lambda = N\lambda_{pr}$, where N is an integer number ≥ 1; the system is designed such that the Bragg condition can be satisfied for N=3. The resulting condition $\lambda > \lambda_{pr}$ is typically advantageous, since in most of the cases this implies that $L_{abs,pr} > L_{abs}$ and thus it is possible to effectively probe the entire volume excited by the pump without a significant signal loss. The polarization of the probe beam is linear and parallel to the y axis (see figure 1(a) for the reference frame), while the polarization of the pump can be either linear (parallel or orthogonal to the probe one) or circular. In special conditions crossed linear polarization for the pump beams is possible [49]. The EUV TG signal can be detected both in forward and backward diffraction in a $\theta_{sig}$ range of ±45°; backward diffraction is realized by introducing a small angle (≈10°) in the z-y plane with respect to the sample surface normal [26]. A signal polarization analysis is also possible [30], while not all combinations of θ and $\theta_{pr}$ are viable; further details on the experimental setup are reported elsewhere [24].

| $\lambda_{pr}$ [nm] | BW [nm] | $T_{peak}$ | Resonance |
|---|---|---|---|
| 20.4 | 1.3 | 0.077 | Co-$M_{2,3}$ |
| 17.7 | 0.6 | 0.093 | Br-$N_{4,5}$ |
| 16.6 | 1.4 | 0.15 | Al-$L_{2,3}$ ; Cu-$M_{2,3}$ Pt-$N_{6,7}$ |
| 13.3 | 0.6 | 0.2 | Fe-$M_1$ ; Kr $M_{4,5}$ ; Ir-$O_1$ |
| 8.34 | 0.1 | 0.014 | Si-$L_1$ ; Gd-$N_{4,5}$ |
| 6.71 | 0.05 | 0.079 | - |

Table I: available settings for the probe beam. The first and second column report, respectively, the central value of $\lambda_{pr}$ and the bandwidth for each setting, the third column is the peak transmission of the delay line (the bandwidth is determined as the FWHM with respect to $T_{peak}$) and the fourth column reports the main absorption edges falling in the respective ranges in $\lambda_{pr}$.

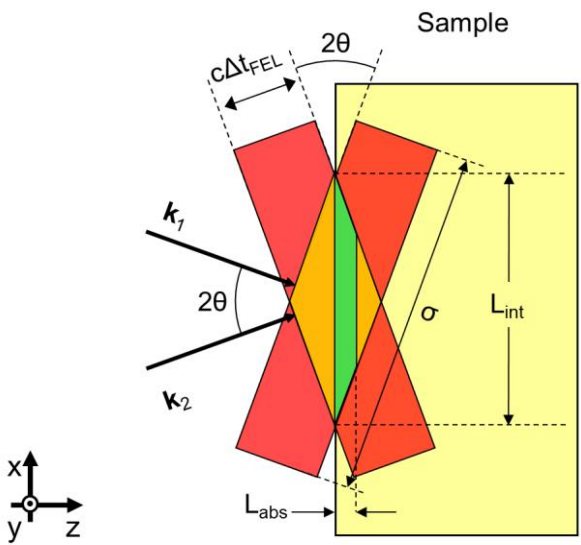

Figure 6: Sketch of the effective interaction region (green area) for two crossed beams with time duration $\Delta t_{FEL}$ and spot size σ (red rectangles), $k_{pump1}$ and $k_{pump2}$ (thick black arrows) are their wavevectors. The reference frame is shown in the bottom left corner.

In light of the low transmission of materials in the EUV spectral range, as illustrated in figure 5(b) for Au and C (common EUV optics coatings), the TIMER instrument is fully based on reflective optics. This choice allowed to have enough transmission to enable EUV TG experiments in a wide wavelength range, that matches the source emission, i.e. ≈ 6.5 – 60 nm. Furthermore, in such a range reflective optics allow for a tangible photon throughput (> 1%) of the beamline, despite the relatively large angles needed to reach values of $\Lambda_{TG}$ as short as ≈10-15 nm (see Eq. 1); we recall here that the reflectivity quickly drops both for small incidence angles and short wavelengths. The drawback of an all reflective setup, is that the two pump beams have a wavefront tilt (WFT) at the sample position equal to 2θ, as shown in figure 6, which inherently limits the effective size of the interaction region, $L_{int}$, i.e. the region where the EUV TG is formed.

Assuming Gaussian pump pulses of equal fluence (F), time duration given by $\Delta t_{FEL}$, and FWHM of the spot σ, even for an infinite spatial extension of the beams (i.e. σ → ∞), the peak to valley amplitude of the TG decreases along the spatial coordinate $x$ due to the delay ($\tau_{pp} = 2x \tan(\theta)/c$; with c the speed of light in vacuum) between two points of the wavefront away from the center of the spot (see figure 6). Therefore, the spatial extension of the interaction region is limited to: $L_{int} \leq c\Delta t_{FEL}/\tan(\theta)$. If $F_{pr}$ and σ$_{pr}$ are, respectively, the fluence and the FWHM spot size of the (Gaussian) probe, the diffraction efficiencies (Eqs. 5 and 6) have to be multiplied by the following factor:

$$\xi = \frac{\int F^2 e^{-8\ln(2)x^2/L_{int}^2} F_{pr} dx dy}{\int F^2 F_{pr} dx dy} = \left(1 + \frac{2\sigma_{pr}^2}{\sigma^2} + \frac{2\sigma_{pr}^2}{L_{int}^2}\right)^{-1/2} \left(1 + \frac{2\sigma_{pr}^2}{\sigma^2}\right)^{-1/2},$$ (Eq. 7)

to account for the finite size of the beams and of $L_{int}$ in such a WFT geometry, where we assumed round spots (same FWHM along x and y), i.e.: $F \propto e^{-4\ln(2)x^2/\sigma^2}$ and $F_{pr} \propto e^{-4\ln(2)x^2/\sigma_{pr}^2}$. Eq. 7 indicates that $\xi$ is significant only when either σ or $L_{int}$ are comparable or smaller than σ$_{pr}$. For example, $\xi < 0.2$ for typical values at the TIMER instrument: $L_{int} \approx$ 15-150 µm and $\sigma_{pr} \approx \sigma \approx$ 30-300 µm (see also Table II and Table III). We emphasize that these values of L$_{int}$ are sufficiently large to prevent the TG signal decay due to the propagation of LA or SAW modes outside the interaction region [50]. Even for a large propagation velocity (say 15 km/s) such a decay is in the order of 1-10 ns, a range typically outside the Δt range accessible by the instrument.

We note that for the relatively large values of $\theta$ needed to achieve short values of $\Lambda_{TG}$, an increase in L$_{int}$ can be obtained by increasing $\Delta t_{FEL}$, if ultrafast time resolution is not required. This is typically not a stringent requirement for detecting the thermoelastic response, even for $\Lambda_{TG} \approx$ 10's nm. On the other hand, when ultrafast time resolution is required, such as for studying electronic and magnetic response with resonant probes, the WFT inherently limits the time resolution. To reach the 1-10 fs level (in principle possible at FERMI) the spot sizes should be reduced to the µm level, with the appropriate attenuation when sample damage does not allow to exploit the total photon flux delivered by the source. A solution to overcome the WFT issue and the trade-off between increasing the size of the interaction region vs preserving the temporal resolution, would be the implementation in the EUV of an approach based on diffractive optics, which is commonly used in the optical regime [51] and may represent the next radical step for the development of EUV/soft x-ray TG. A similar scheme (Talbot carpet) has been attempted in the x-ray with optical probe [52].

Equations 4-7 are useful tools to estimate the TG efficiency due to the thermoelastic response, which in the EUV range exploitable by the TIMER instrument we roughly estimate to be in the $\eta_{B,F} \approx 10^{-7} - 10^{-10}$ range. Such equations can also be applied to other spectral ranges, and are particularly useful when the absorption lengths are comparable with other experimental length-scales such as $\Lambda_{TG}$, $\lambda_{pr}$ and d. We notice that Eqs. 5 and 7 arise only from finite absorption and experimental geometry considerations and, therefore, are not restricted to the context of thermoelasticity.

## 4 Experimental results

Figure 7 displays representative EUV TG data collected in forward diffraction geometry from different samples, namely: amorphous Si$_3$N$_4$ (panels (a) and (b) [4,14]), the amorphous metallic alloy Zr$_{65}$Cu$_{27.5}$Al$_{7.5}$ (panels (c) and (d)) and the crystalline atomic metal Co (panel (e)); the last two samples were grown on a thin Si$_3$N$_4$ membrane. The total thickness of all samples was sufficiently small (< 150 nm) to permit tangible transmission of the signal beam. Data are plotted in terms of the estimated, time dependent, diffraction efficiency ($\eta_F^*(\Delta t)$) as obtained from the nominal conversion of detector counts into photons (the detector is a CCD camera, Princeton Instruments MTE 2048B) and the estimated transmission of the beamline including filters. These are placed both along the beam path and in front of the detector, to remove spurious components in the FEL emission and background light. The beamline and filter transmission are the largest source of uncertainty, since deviations from the nominal values as large as 50% were observed due to progressive damaging and contamination from residual gas pressure in the vacuum vessels. The acquisition time for a signal waveform ranges from about one to a few hours. Experimental data have been collected in different conditions, summarized in Table II. Since data were mostly collected in the commissioning and initial operation phase of the instrument, it was not yet possible to evaluate all the beamline and FEL parameters that were later on deemed necessary from experience. Therefore, the comparison between the magnitude of $\eta_F^*(\Delta t)$ and the values expected from the above equations should be kept on a qualitative level.

| Fig. | Material | d [nm] | $\Lambda_{TG}$ [nm] | F [mJ/cm$^2$] | $\lambda_{pr}$ [nm] | L$_{abs}$ [nm] | L$_{abs,pr}$ [nm] | $\xi$ | $\Delta Q_z/k_{pr}$ |
|---|---|---|---|---|---|---|---|---|---|
| 7(a) | SiN | 100 | 28 | 22 | 13.3 | 118 | 118 | 0.19 | 0.08 |
| 7(b) | SiN | 100 | 84 | 3.9 | 13.3 | 17 | 118 | 0.19 | < 10$^{-3}$ |
| 7(c) | Zr$_{65}$Cu$_{27.5}$Al$_{7.5}$/SiN | 39/100 | 24 | 7.3 | 13.3 | 12 | 68 | 0.004 | < 10$^{-3}$ |
| 7(d) | Zr$_{65}$Cu$_{27.5}$Al$_{7.5}$/SiN | 39/100 | 84 | 0.9 | 13.3 | 12 | 68 | 0.1 | < 10$^{-3}$ |
| 7(e) | Co/SiN | 20/100 | 44 | 15 | 20.8 | 12 | 12 | 0.1 | 0.01 |

*Table II: Summary of experimental and sample (membranes) parameters for data shown in figure 7.*

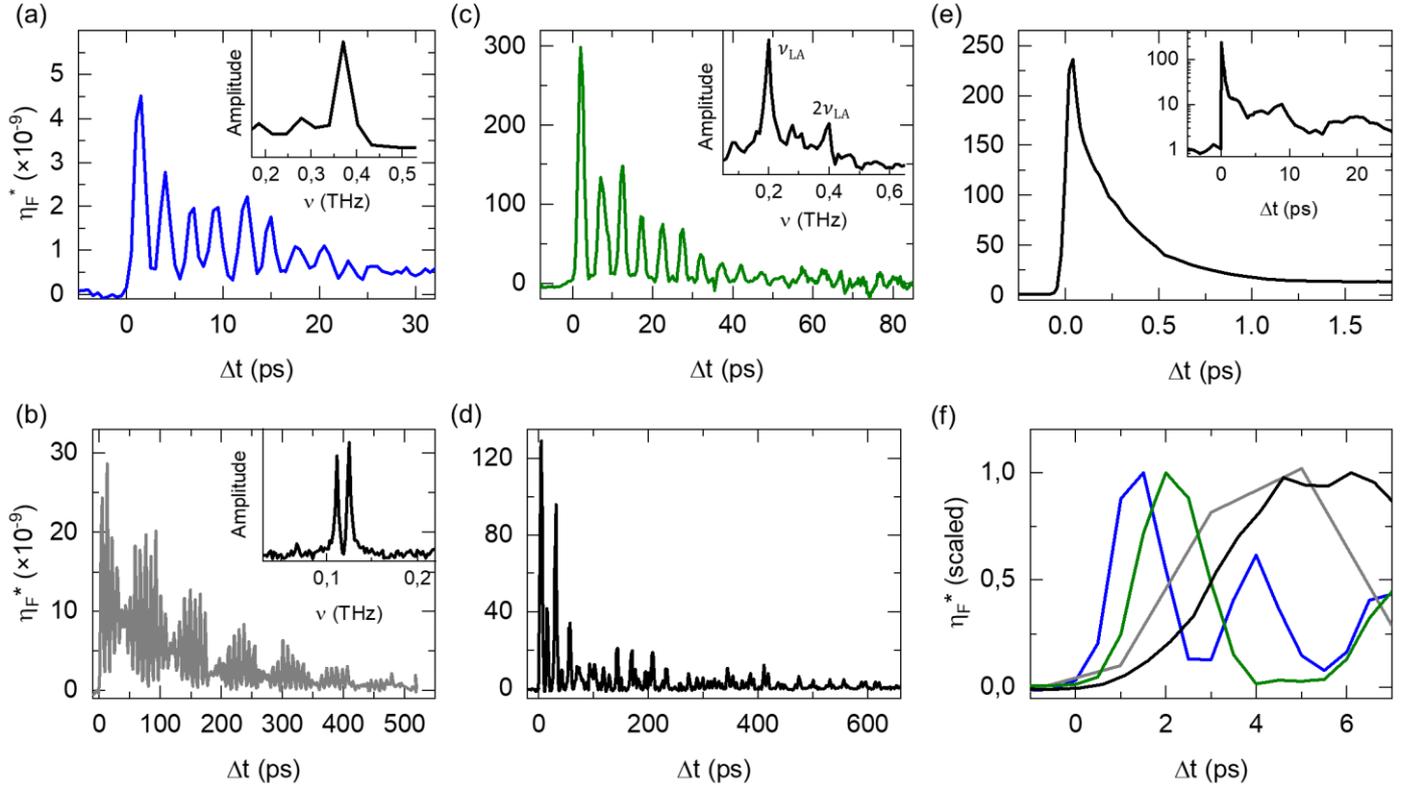

*Figure 7: EUV TG signal in forward diffraction as a function of Δt from different samples, namely: amorphous Si$_3$N$_4$ at $\Lambda_{TG}$ =28 nm (panel (a)) and at $\Lambda_{TG}$ =84 nm (panel (b)) [4,14], Zr$_{65}$Cu$_{27.5}$Al$_{7.5}$ at $\Lambda_{TG}$ =24 nm (panel (c)) and at $\Lambda_{TG}$ =84 nm (panel (d)) [53], and Co at $\Lambda_{TG}$ =44 (panel (e)). The inset in the latter panel shows, in a semi logarithmic scale, the signal waveform on a longer range in Δt; we vertically offset $\eta_F^*(\Delta t)$ by a factor $10^{-9}$ to fit in the logarithmic vertical scale. See Table II and main text for sample parameters and experimental conditions. The insets in panels (a)-(c) show the Fourier spectrum of signal waveforms, after subtracting the slow thermal decay, the labels $\nu_{LA}$ and $2\nu_{LA}$ in the inset of panel (c) indicate the single and double frequency peaks, both due to the same LA mode (see text for discussion). Panel (f) displays data from panels (a)-(d), keeping the same color code, in a narrower range in Δt and with the amplitude scaled to the maximum of the first signal oscillation.*

In general, the efficiency of the EUV TG signals in figure 7 matches the aforementioned order of magnitude estimate ($\eta_{B,F} \approx 10^{-7} - 10^{-10}$). The lower intensity signal from the insulating sample Si$_3$N$_4$ (figure 7(a)- 7(b)) with respect to the metallic alloy (figure 7(c)- 7(d)) can be explained in terms of the larger value of $\alpha_v$ expected for metals. The data in figure 7(c) is obtained with a much smaller value of $\xi$ due to an exceptionally large dimension of the probe beam in that specific experiment. Nevertheless, its comparison with the signal in figure 7(d) demonstrates how a small $\xi$ can be compensated with a larger fluence F. Finally, the signal waveform in figure 7(a) shows how a tangible EUV TG signal can be detected even for $\Delta Q_z/k_{pr} \approx 0.08$ (see Table II). As discussed referring to figure 4(a), this is possible because of the strong EUV absorption and, indeed, the TG signal would be essentially negligible in a non-absorbing sample.

The Δt-dependencies of the EUV TG signal waveforms showed in figure 7 reflect the time dependence of the factor $|\Delta n(\lambda_{pr})|^2$ in Eqs.3 and 5. For panels (a)-(d) it consists in a slow decay of $\eta_F^*(\Delta t)$, which can be ascribed to thermal transport over a characteristic distance $L_{th} = \Lambda_{TG}/\pi$ [54], modulated by oscillations due to propagation of acoustic phonons, i.e.:

$$\eta_F^*(\Delta t) \propto \left| A_{th} e^{-\Delta t/\tau_{th}} - \sum_i A_i \cos(2\pi\nu_i \Delta t) e^{-\Delta t/\tau_i} \right|^2, \qquad \text{(Eq. 8)}$$

where $\tau_{th}$ is the thermal decay constant, while $\nu_i$ and $\tau_i$ are, respectively, the phonon frequency and lifetime of the $i^{th}$ phonon mode in the waveform [4,14,26]. Data in figures 7(a)-7(d) illustrate how there is no qualitative difference between the signal from insulating (7(a) and 7(b)) and metallic samples (7(c)-7(d)), as expected according to the previous discussion. This makes EUV TG a useful tool to study the thermoelastic response in bulk metals, where optical TG experiments are hardly feasible, with the additional possibility to access the nanoscale with ultrafast time resolution; a detailed analysis of the thermoelastic response from atomic metals will be the subject of a separate manuscript [55].

Data in figure 7(a) and 7(b) were collected at different values of $\Lambda_{TG}$ from nominally the same sample, i.e. a 100 nm thick membrane of amorphous Si$_3$N$_4$, and show the role of the ratio $d/\Lambda_{TG}$. As discussed in Ref. [14,27,56], for $2\pi d/\Lambda_{TG} > 10$ the thermoelastic response of a membrane can be approximated by that of the bulk ($d \to \infty$), which, in amorphous samples and in the employed geometry, results in a single LA phonon mode [57]. This is the situation of figure 7(a), where $2\pi\,d/\Lambda_{TG} \approx 22$ and the Fourier spectrum of the signal waveform (shown in the inset) features a single peak matching the expected LA frequency at the employed value of $\Lambda_{TG}$. Conversely, in figure 7(b) $2\pi\,d/\Lambda_{TG} \approx 7.5$ and the signal exhibits a clear beating. Such beating arises from two modes with similar frequency (see inset for the Fourier spectrum), both close to the LA frequency and with a frequency splitting determined by the coupling between LA and transverse acoustic (TA) modes. This coupling permits to gain information on nanoscale phonons of both LA and TA nature. A detailed discussion on the effects of the parameter $2\pi\,d/\Lambda_{TG}$ in amorphous $Si_3N_4$ is reported in Refs. [14,27]. The same phenomenon is also responsible for the complex waveform in figure 7(d) compared to the one in figure 7(c). In this case the sample is a bilayer consisting of a 39 nm thick amorphous metallic alloy ($Zr_{65}Cu_{27.5}Al_{7.5}$) deposited on a 100 nm membrane of amorphous $Si_3N_4$. Indeed, for $2\pi\,d/\Lambda_{TG} \approx 2.9$ (figure 7(d)), the signal shows a complex beating pattern, most likely arising from coupling between LA and TA modes in the bi-layer system, while for $2\pi\,d/\Lambda_{TG} \approx 10$ (figure 7(c)) the signal waveform essentially contains a single oscillation frequency, compatible with the LA mode. Note that in the latter case a component at twice the frequency can be faintly perceived for $\Delta t > 40$ ps and is evident in Fourier spectrum (shown in the inset). This is not related to an additional phonon mode, but to the decay of the thermal response occurring before that of the oscillatory component. Indeed, in case of a single phonon mode, when the thermal decay is over ($e^{-\Delta t/\tau_{th}} \to 0$) Eq. 8 reduces to $\eta_F^*(\Delta t) \propto |\cos(2\pi\nu_i \Delta t)|^2$, which shows oscillations at $2\nu_i$.

The data in figure 7(a)- 7(d) were collected with EUV photon energies far from any core resonances of the materials. Therefore, we do not observe any noticeable signal from the initial electronic population grating (see sketch in figure 2(a)). Conversely, data in figure 7(e) were acquired with the probe tuned to an electronic core-hole transition (Co M-edge; $\lambda_{pr} = 20.8$ mn) and illustrate how this enables to detect the electronic response in addition to the thermoelastic one. The electronic response signature is the prominent feature at $\Delta t = 0$, which is approximatively 50 times brighter than the amplitude of the first thermoelastic oscillation (i.e. the EUV TG signal in the 5-25 ps range, as shown in the inset) and consists in a peak with a rise time in the order of $\Delta t_{FEL}$, compatible with the experimental time resolution (< 100 fs), and a decay time of about 250 fs, in the order of typical electron-phonon relaxation timescales (see sketch in figure 2(a)). In order to better highlight such a substantially different response, we show in figure 7(f) the signal waveforms displayed in figure 7(a)-7(d) on a shorter $\Delta t$ range. One can readily notice how the EUV TG signal slowly rises (slower with respect to the $\Delta t=0$ peak in figure 7(e)), following a sinusoidal waveform and reaching a maximum corresponding to half of the acoustic period; note that for a given material such a maximum is achieved at longer $\Delta t$ since the phonon period scales as $\Lambda_{TG}$.

Figure 8 shows some EUV TG data collected in backward diffraction geometry under different experimental conditions, except for panel (a) that reports a signal waveform collected in forward diffraction. Experimental and sample details are summarized in Table III. Data are plotted in terms of the estimated, time dependent diffraction efficiency ($\eta_B^*(\Delta t)$), obtained as explained above for $\eta_F^*(\Delta t)$ and under the same considerations. The acquisition time for a signal waveform is similar to data acquired in forward diffraction, i.e. from about one hour to a few hours.

| Fig. | Material | d [nm] | $\Lambda_{TG}$ [nm] | F [mJ/cm$^2$] | $\lambda_{pr}$ [nm] | $L_{abs}$ [nm] | $L_{abs,pr}$ [nm] | $\xi$ | $\Delta Q_z/k_{pr}$ |
|---|---|---|---|---|---|---|---|---|---|
| 8(a) | $SiO_2$ | 100 | 84 | 1.8 | 13.3 | 17 | 84 | 0.1 | < 10$^{-3}$ |

| Fig. | Material | d [nm] | $\Lambda_{TG}$ [nm] | F [mJ/cm$^2$] | $\lambda_{pr}$ [nm] | R [10$^{-4}$] | $\xi$ |
|---|---|---|---|---|---|---|---|
| 8(b) | $SiO_2$ | 100 | 84 | 1.8 | 13.3 | 0.64 | 0.1 |
| 8(c) | $SiO_2$ | Bulk | 84 | 0.9 | 13.3 | 1.5 | 0.14 |
| 8(d) | $YBa_2Cu_3O_7$ | Bulk | 69 | 2.4 | 16.5 | 0.007 | 0.074 |
| 8(e) | $TiO_2:Ta_2O_5$ | 500 | 56 | 0.35 | 13.3 | 8.2 | 0.14 |
| 8(f) | $SrTiO_3$ | Bulk | 84 | 0.65 | 13.3 | 6.6 | 0.1 |

*Table III: Summary of experimental and sample parameters for data shown in figure 8; to calculate R we assumed a roughness equal to 1 nm (root-mean-square) for SiO$_2$ membrane, which is typical of this type of samples, 3.5 nm for YBa$_2$Cu$_3$O$_7$, according to a characterization measurement done with an atomic force microscopy, while for the other samples we assumed a negligible effect, which is likely in highly polished bulk materials.*

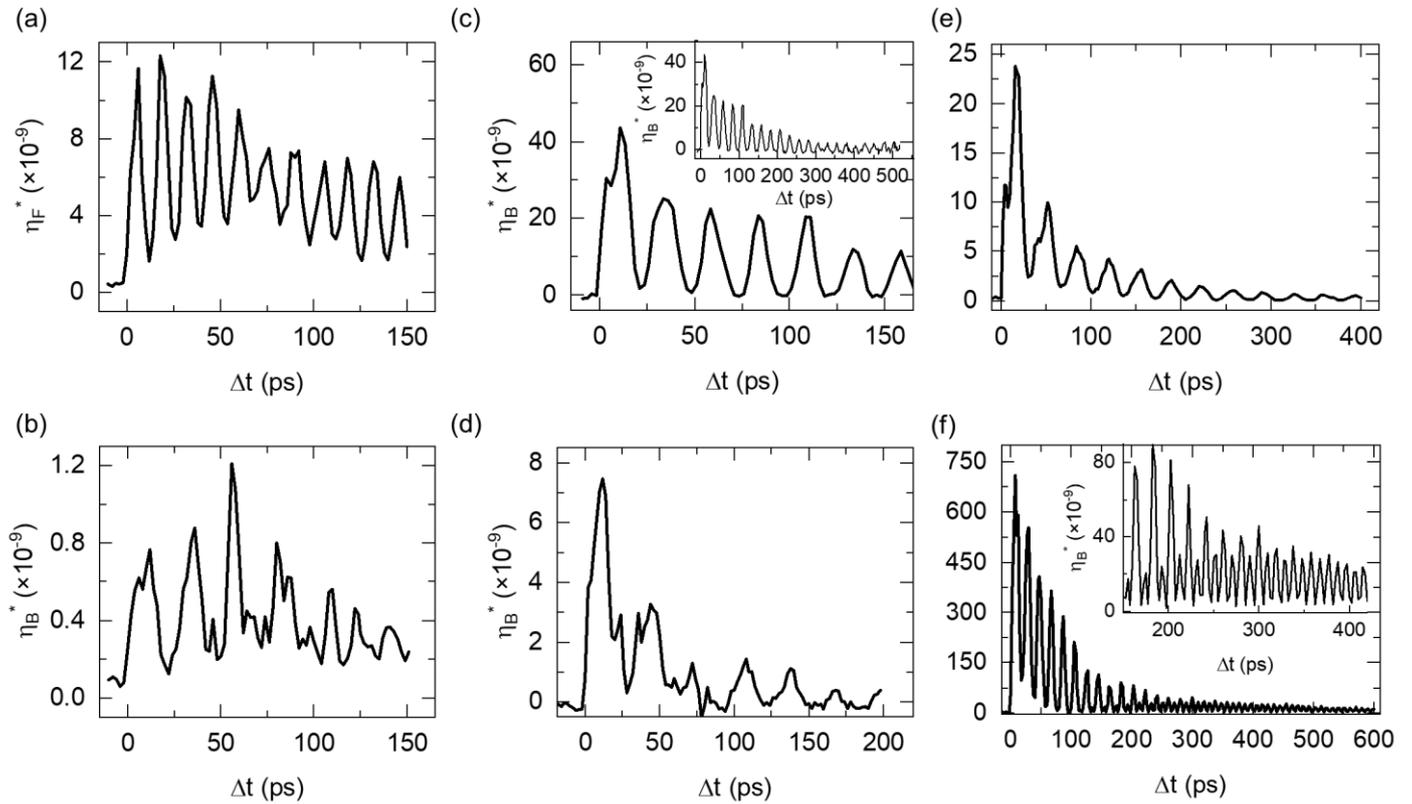

*Figure 8: EUV TG signal from different samples, namely: amorphous SiO2 at $\Lambda_{TG}$ =84 nm, collected in both forward (a) and backward diffraction, (b) and (c); see text for further details. Inset of panel (c) shows the full decay of the SAW. (d) polycrystalline YBa2Cu3O7 sample at $\Lambda_{TG}$ =69 nm. (e) EUV TG signal in backward diffraction from a TiO2:Ta2O5 mirror used in the Virgo apparatus [58], in this case $\Lambda_{TG}$ =56 nm. (f) single crystalline SrTiO3 at $\Lambda_{TG}$ =84 nm; the inset shows the decay of the double frequency response for Δt > 250 ps) [26].*

Figures 8(a) and 8(b) compare EUV TG signal waveforms at $\Lambda_{TG}$ = 84 nm, collected respectively in forward and backward diffraction geometry, from the same 100 nm thick membrane of amorphous $SiO_2$. These data were collected one right after the other, in the same experimental conditions. It can be readily noticed how the magnitude of $\eta_F^*(\Delta t)$ and $\eta_B^*(\Delta t)$ is of similar order, while the oscillation frequency substantially changes (note the same horizontal scale), matching accordingly the expected frequency of SAW and LA modes. This is due to the fact that EUV TG in backward diffraction is mainly sensitive to surface displacement, which dynamics is dominated by SAWs, while in forward diffraction the signal is dominated by LA modes (bulk density modulations). These plots demonstrate how EUV TG experiments performed in both forward and backward diffraction geometry can be exploited to detect bulk and surface phonons from the same sample with comparable efficiency. Another advantage of backward diffraction in the regime of strong absorption is to remove the need of thin samples. In many practical cases the surface of bulk samples can be polished to reach a roughness better than membranes, resulting in higher reflectivity (see figure 3(a)) and thus larger signal levels. An example of this situation is evident when comparing the $SiO_2$ membrane signal of figure 8(b) with figure 8(c), where the EUV TG signal was collected from an ultra-polished surface of bulk $SiO_2$. It can be readily seen that in this case the signal level is larger and the overall quality of the measurement superior. Such conditions allow to detect low signal levels and thus to reliably monitor, e.g., the signal decay, as shown in the panel inset. Additionally, many samples cannot be fabricated in the form of sub-μm membranes, as for instance polycrystalline YBa$_2$Cu$_3$O$_7$ (figure 8(d)), or one may need to study a specific sample designed for other purposes, as it was the case of the titania-doped-tantala mirrors (TiO$_2$:Ta$_2$O$_5$, concentration ratio Ti/Ta = 0.21, annealed in air for 10 hours at 500 °C) used in the Virgo apparatus (figure 8(e)) [58] and produced by the Laboratoire des Matériaux Avancés of the CNRS [59] as the standard for the LIGO-Virgo collaboration.

As a time-domain approach, EUV TG can detect long time dynamics, such as phonon decays, better than frequency resolved approaches. This capability was used, for example, to determine the lifetime of high frequency SAWs in SrTiO$_3$, as discussed in Ref. [26] and shown in figure 8(f). The transition from a signal waveform oscillating at Ω, which mainly due to the thermal decay modulated by the SAW, to a 2Ω signal, which appears when the thermal decay is completely over (Δt > 250 ps) and reflects the pure SAW response, can be clearly seen in the inset. Since LA, TA and SA waves show linear dispersions as a function of the inverse TG period, i.e. $\Omega \propto \Lambda_{TG}^{-1}$, the use of EUV TG allows to probe a frequency range largely exceeding the one accessible by optical methods. Furthermore, in the context of SAWs, EUV TG represents a unique tool, since neither thermal neutrons nor x-rays can effectively detect such surface modes (despite a few attempts [60]), because of their large penetration depth, while both electron and He atoms spectroscopies have limited resolution and dynamic range.

## 5  Conclusion and perspectives

In this manuscript we summarize the experience acquired from EUV TG experiments aimed at studying the thermoelastic response of solids in the sub-100 nm length-scale, that were carried out at the TIMER instrument of the FERMI FEL in the last 5 years. We focus on the peculiarities related to the use of EUV light both for TG excitation and probe, in both forward and backward diffraction geometry. In the first case these peculiarities are related mainly to the strong absorption of EUV photons in condensed matter, while in the latter they are related to the behavior of EUV reflectivity. Furthermore, we also discuss the effects of the finite beam size and of the interaction region, in the context of the employed experimental geometry, which introduces a tangible wavefront tilt between the crossing pump pulses. All these considerations are useful to evaluate the efficiency of EUV TG experiments, but since they essentially arise from general aspects of the TG process, can also be applied to optical TG and can be relevant, e.g., when either the pump or probe beam (or both) are strongly absorbed by the sample.

We show some selected EUV TG data to illustrate the reliable capability of EUV TG of probing bulk, confined and surface phonon modes, as well as thermal transport, in the sub-100 nm scale, from both insulating and metallic samples, without any need of nanostructuring or contacting the sample. Such a capability can be profitably exploited in several fields, such as high frequency surface waves, structural dynamics in disordered systems or phonon engineering in nanostructures. Since data were mostly collected in the commissioning and initial operation phase of the instrument, it was not possible to store all the beamline and FEL parameters that were deemed necessary with experience. This limits the comparison between the experimental and expected signal to a qualitative level, which is nevertheless quite satisfactory. Nowadays, the control over the experimental parameters has been improved by better diagnostics and procedures, while EUV optics able to provide tighter focusing of the probe (to reach values of $\xi \approx 1$) are under evaluation. Finally, we mention the capability to exploit electronic core-hole transitions to probe the ultrafast electronic or magnetic responses, in addition to the thermoelastic one, with the added value of element specificity. This is an example of application which was not anticipated at the beginning of the project. Similarly, since optical TG is widely applied to disparate fields, we expect that other types of experiments exploiting the capability to generate sub-100 nm TGs, may be devised in the short-term future.

A couple of considerations are worth mentioning in the context of future developments for EUV/x-ray TG spectroscopies. Firstly, the aim of pushing TG to shorter spatial scales could be achieved by exploiting new mirror coating technologies that would allow to push the TIMER instrument towards shorter wavelengths. This would also be beneficial for spectroscopy, since it could potentially open the investigation of relevant core-hole resonances such as the L-edges of 3d transition metals around 1-2 nm or the K-edges of C, N and O around 2-4 nm that are fundamental for the investigation of organic compounds. Changing completely the experimental paradigm, however, similar photon energy ranges could be accessed with an improved phase-mask technology. Moving from fully reflective to diffractive setups would mitigate, if not completely suppress, the problematics associated with the wavefront tilt and thus preserve the temporal resolution required to investigate dynamics occurring within the initial steps of electronic excitation. Similarly, mastering diffractive setups is of great advantage for x-ray-based TG, where the strong reflectivity drop prevents the exploitation of reflective geometries. Wavelengths on the order of few Å could potentially enable the exchanged momentum to reach the Brillouin zone and beyond, and thus investigate collective excitations such as charge density waves. Moreover, the limitations in the signal intensity due to strong absorption are expected to be strongly mitigated, at the price of much stricter constraints dictated by phase matching. In addition, the extended access to deeper and sharper core transitions will improve element specific spectroscopy. As a drawback, the decrease in reflectivity will hamper the capability of performing TG in backward diffraction, thus limiting x-ray TG to bulk spectroscopy.

The second consideration regards the possibility of extending EUV TG to a broader community, nowadays limited by the very competitive access to FEL facilities. New HHG sources might be soon capable of generating TGs at least at the longer EUV wavelengths. This would be sufficient to reach periodicities around 100 nm on table-top setups, with the additional benefit of high repetition rate improving drastically the signal-to-noise ratio and strongly reducing acquisition times. With the same goal in mind, and especially for the investigation of slower thermoelastic responses, it would be interesting to consider exploring synchrotron facilities operating in time-resolved mode. Indeed, they could provide sufficiently short pulses (few ps), very high (GHz) repetition rates, translating in an average power comparable to the one of FELs, to a very broad and experienced audience.

## 6 Acknowledgments:

The authors acknowledge the Laboratoire des Matériaux Avancés (LMA-CNRS) for providing the mirror with the standard coating for the LIGO-Virgo collaboration.
A.A.M, C.A.O, J.L., R.C., and K.A.N. received support from the Department of Energy, Office of Science, Office of Basic Energy Sciences, under Award Number DE-SC0019126
F.Capor. acknowledges financial support from The Netherlands Organization for Scientific Research (NWO) (Grant Number 680-91-13).
G.M. acknowledges support from the research project GLASS@EXTREMES financed by the Cariparo foundation (2021).

# 8 Appendix

**Derivation of Eq. 3 and 5**

We start with a thin grating of the complex refractive index $\Delta n(\lambda_{pr}) \cos qx$ and thickness $d$.

The electric field amplitude in the incident probe beam (polarized along y) is given by

$$E_{pr} = e^{-ik \sin \theta_{pr} x - ik \cos \theta_{pr} z},$$

Where $k = 2\pi n(\lambda_{pr})/\lambda_{pr}$ and $\theta_{pr}$ are the wavevector magnitude and the incidence angle of the probe beam *in the medium*. At the output of the thin grating, positioned at $z = z_1$, the electric field is given by

$$E = e^{-ik \sin \theta_{pr} x - ik \cos \theta_{pr} z_1 - ikd\Delta n(\lambda_{pr}) \cos qx/\cos \theta}. \tag{Eq. S1}$$

Assuming that $kd\Delta n(\lambda_{pr})$ is infinitesimally small, we expand $e^{-ikd\Delta n(\lambda_{pr}) \cos qx/\cos \theta}$ in a Taylor series and retain only the first order terms, obtaining:

$$E = e^{-ik \sin \theta_{pr} x - ik \cos \theta_{pr} z_1} \left[1 - \frac{ik\Delta n(\lambda_{pr})d}{2 \cos \theta_{pr}} \left(e^{iqx} + e^{-iqx}\right)\right] \tag{Eq. S2}$$

Consequently, the diffracted field at $z > z_1$ is given by a superposition of three plane waves with $x$ components of the wavevector $k \sin \theta_{pr}$ (0$^{th}$ diffraction order), $k \sin \theta_{pr} + q$ and $k \sin \theta_{pr} - q$ ($\pm 1$ diffraction orders). Let us consider one of these diffraction orders propagating at an angle $\theta_{sig}$ such that $k \sin \theta_{sig} = k \sin \theta_{pr} + q$. The electric field in this diffraction order is given by:

$$E_{sig} = -\frac{ik\Delta n(\lambda_{pr})d}{2\cos\theta_{pr}} e^{-i\Delta Q_z z_1} e^{-ik\sin\theta_{pr} x - ik\cos\theta_{sig} z}, \tag{Eq. S3}$$

where $\Delta Q_z = k(\cos\theta_{pr} - \cos\theta_{sig})$.

Considering that the incoming probe field $E_{pr}$ and the outgoing signal field $E_{sig}$ have different widths, the diffraction efficiency of a thin grating is given by

$$\eta = \frac{\cos\theta_{sig}}{\cos\theta_{pr}} \frac{|E_{sig}|^2}{|E_{pr}|^2} = \frac{\cos\theta_{sig}}{\cos\theta_{pr}} |\Delta n(\lambda_{pr})|^2 \left[\frac{\pi n(\lambda_{pr})d}{\lambda_{pr} \cos \theta_{pr}}\right]^2, \tag{Eq. S4}$$

where we have substituted $k = 2\pi n(\lambda_{pr})/\lambda_{pr}$. Note that Eq. S4 corresponds to the limit of Eq. 3 for $d \to 0$.

Let us now consider diffraction by a TG in a thick slab extending from $z_1 = 0$ to $z_1 = d$, with the refractive index modulation $\Delta n(\lambda_{pr})$ in the TG having a depth profile determined by the absorption length of the pump pulse intensity $L_{abs} \cos \theta$,

$$\Delta n(\lambda_{pr}) = \Delta n_{z=0}(\lambda_{pr}) e^{-z/L_{abs} \cos \theta}.$$

We represent our thick slab as consisting of an infinite number of thin gratings, whose diffracted fields add up. If the absorption of probe light in the slab is disregarded, then in order to calculate the diffracted field, we need to integrate Eq. S3 over *z1* from 0 to *d*. This would result in Eq. 3.

If the absorption is not negligible, we need to introduce additional factors accounting for the attenuation of the probe field before it comes to a given thin grating located at $z_1$ ($e^{-z_1/2L_{abs,pr} \cos \theta_{pr}}$) as well as for the attenuation of the diffracted field before it leaves the slab, ($e^{-(d-z_1)/2L_{abs,pr} \cos \theta_{sig}}$), i.e.:

$$E_{sig} = -\frac{ik\Delta n(\lambda_{pr})d}{2\cos\theta_{pr}} e^{-ik\sin\theta_{pr} x - ik\cos\theta_{sig} z} \int_0^d e^{-i\Delta Q_z z_1} e^{-z_1/L_{abs} \cos \theta} e^{-z_1/2L_{abs,pr} \cos \theta_{pr}} e^{-(d-z_1)/2L_{abs,pr} \cos \theta_{sig}} dz_1 \tag{Eq. S5}$$

The integration yields

$$E_{sig} = \frac{k\Delta n(\lambda_{pr})d}{2\cos\theta_{pr}} e^{-ik\sin\theta_{sig} x - ik\cos\theta_{sig} z} \frac{e^{-d/(2L_{abs,pr}\cos\theta_{pr})} \left(e^{-i\Delta Q_z d - d/2L^*} - 1\right)}{\Delta Q_z d - id/2L^*}, \tag{Eq. S6}$$

Where $L^* = \left((L_{abs} \cos \theta/2)^{-1} + (L_{abs,pr} \cos \theta_{pr})^{-1} - (L_{abs,pr} \cos \theta_{sig})^{-1}\right)^{-1}$. The final result is then:

$$\eta_F = \frac{\cos \theta_{sig}}{\sin \theta_{pr}} \frac{|E_{sig}|^2}{|E_{pr}|^2} = \frac{\cos \theta_{sig}}{\sin \theta_{pr}} |\Delta n(\lambda_{pr})|^2 \left[\pi d n(\lambda_{pr})/\lambda_{pr} \cos \theta_{pr}\right]^2 \frac{e^{-d/L^*} - 2\cos(\Delta Q_z d)e^{-d/2L^*} + 1}{(d/2L^*)^2 + (\Delta Q_z d)^2} e^{-d/L_{abs,pr} \cos \theta_{sig}}. \tag{Eq. 5}$$